\documentstyle[12pt]{article}
\begin{document}

\title{Hamiltonian Formulation of Two Body Problem in Wheeler--Feynman
electrodynamics}
\author{I.N.~Nikitin\\   \\
Institute for High Energy Physics -- Protvino, Russia\\
E-mail address:\ \ nikitin\_{$\,$}i@mx.ihep.su}
\date{ }
\maketitle

\vspace{2cm}
{\bf Summary. } --  A problem of electromagnetic interaction of two charged
relativistic particles is considered in Wheeler--Feynman approach to the
classical electrodynamics. We formulate this theory in such a way that
its Hamiltonian description becomes available. This description is a kind
of constrained Hamiltonian mechanics. Hamiltonian equations obtained have
simpler form than Lagrangian ones.

\vspace{0.5cm}

PACS 03.50.De -- Maxwell theory: general mathematical aspects

PACS 03.20 -- Classical mechanics of discrete systems: general mathematical
aspects

\vspace{1.5cm}

\newcommand{\nn}{\nonumber}
\newcommand{\sini}
{\begin{eqnarray}
S&=&\int d\tau\ (-m_{1}\sqrt{\dot x^{2}}-m_{2}\sqrt{\dot y^{2}})
\ \nn\\&&-\ e_{1}e_{2}\int d\tau_{1}d\tau_{2}\ {{dx_{\mu}}\over{d\tau_{1}}}
{{dy^{\mu}}\over{d\tau_{2}}}\ \delta((x(\tau_{1})-y(\tau_{2}))^{2})
\label{sini}
\end{eqnarray}}
\newcommand{\dt}{\begin{array}[b]{c}{\bf .}\\ \ \end{array} }
\newcommand{\dto}{^{^{\bf .}}}
\newcommand{\stptm}
{\begin{eqnarray}
S&=&\int d\tau\ \Bigl(-m_{1}\sqrt{\dot x^{2}}-m_{2}\sqrt{\dot y^{2}}\nn\\
&&- {{e_{1}e_{2}}\over{2}}\Biggl( {{\dot x(\tau)y(\tau^{+}(\tau))\dto}\over
{|\dot x(\tau)\Bigl(x(\tau)-y(\tau^{+}(\tau))\Bigr)|}} +
{{\dot x(\tau)y(\tau^{-}(\tau))\dto}\over
{|\dot x(\tau)\Bigl(x(\tau)-y(\tau^{-}(\tau))\Bigr)|}}
\Biggr)\ \Bigr)\label{stptm}
\end{eqnarray}}
\newcommand{\yx}
{\begin{equation}
y^{+}(\tau)=x^{+}(\tau),\quad y^{-}(\tau)=x^{-}(\tau +1)\label{yx}
\end{equation}}
\newcommand{\sr}
{\begin{equation}
S=\int d\tau \Bigl( -m_{1}\sqrt{\dot x^{+}\dot x^{-}}
-m_{2}\sqrt{\dot x^{+}\dot x^{-}_{a}} - e_{1}e_{2}\Biggl(
{{\dot x^{-}}\over{x^{-}_{a}-x^{-}}} + {{\dot x^{+}}\over{x^{+}_{a}-x^{+}}}
\Biggr)\ \Bigr) \label{sr}
\end{equation}}
\newcommand{\eqlag}
{\begin{eqnarray}
\Biggl(m_{1}\sqrt{{{\dot x^{-}}\over{\dot x^{+}}}} +
m_{2}\sqrt{{{\dot x^{-}_{a}}\over{\dot x^{+}}}}\ \Biggr)\dt&=&
2e_{1}e_{2} \Biggl({{\dot x^{+}_{a}}\over{(x^{+}-x^{+}_{a})^{2}}} -
{{\dot x^{+}_{r}}\over{(x^{+}-x^{+}_{r})^{2}}}\Biggr) \label{eqlag}\\
\Biggl(m_{1}\sqrt{{{\dot x^{+}}\over{\dot x^{-}}}} +
m_{2}\sqrt{{{\dot x^{+}_{r}}\over{\dot x^{-}}}}\ \Biggr)\dt&=&
2e_{1}e_{2} \Biggl({{\dot x^{-}_{a}}\over{(x^{-}-x^{-}_{a})^{2}}} -
{{\dot x^{-}_{r}}\over{(x^{-}-x^{-}_{r})^{2}}}\Biggr) \nn
\end{eqnarray}}
\newcommand{\eqmin}
{\begin{eqnarray}
&&m_{1}{{d}\over{dx_{0}}}{{v_{x}}\over{\sqrt{1-v_{x}^{2}}}}=e_{1}E_{x},\quad
m_{2}{{d}\over{dx_{0}}}{{v_{y}}\over{\sqrt{1-v_{y}^{2}}}}=-e_{2}E_{y}
\label{eqmin}\\
&&E_{x}={{e_{2}}\over{2}}\Biggl({{1+v_{y}^{r}}\over{1-v_{y}^{r}}}\
{{1}\over{(x_{1}-y_{1}^{r})^{2}}} + {{1-v_{y}^{a}}\over{1+v_{y}^{a}}}
{{1}\over{(x_{1}-y_{1}^{a})^{2}}}\Biggr)\nn\\
&&E_{y}={{e_{1}}\over{2}}\Biggl({{1-v_{x}^{r}}\over{1+v_{x}^{r}}}\
{{1}\over{(y_{1}-x_{1}^{r})^{2}}} + {{1+v_{x}^{a}}\over{1-v_{x}^{a}}}
{{1}\over{(y_{1}-x_{1}^{a})^{2}}}\Biggr) \nn
\end{eqnarray}}
\newcommand{\eqstar}
{\begin{eqnarray}
m_{1}\Biggl(\sqrt{{{\dot x^{+}}\over{\dot x^{-}}}}\Biggr)\dt&=&
2e_{1}e_{2}\Biggl({{\dot x^{+}\dot x^{+}_{r}}\over{\dot x^{-}(x^{+}-x^{+}_{r}
)^{2}}} + {{\dot x^{-}_{a}}\over{(x^{-}-x^{-}_{a})^{2}}}\Biggr)\label{eqstar}\\
m_{2}\Biggl(\sqrt{{{\dot x^{-}_{a}}\over{\dot x^{+}}}}\Biggr)\dt&=&
2e_{1}e_{2}\Biggl({{\dot x^{-}\dot x^{-}_{a}}\over{\dot x^{+}(x^{-}-x^{-}_{a}
)^{2}}} + {{\dot x^{+}_{a}}\over{(x^{+}-x^{+}_{a})^{2}}}\Biggr)\nn
\end{eqnarray}}
\newcommand{\eqls}
{\begin{equation}
L_{1}=R_{1}+{{H}\over{\dot x^{-}}},\quad L_{2}=R_{2}+{{H}\over{\dot x^{+}}}
\label{eqls}
\end{equation}}
\newcommand{\pro}
{$$(\ref{eqlag})\ \Longleftrightarrow\
\left\{ \begin{array}{l}
-{{\dot x^{-}}\over{\dot x^{+}}} L_{1} + L_{2}=R^{+}\quad
\mbox{($R^{\pm}$ -- the right hand sides of equations (\ref{eqlag}))}\\
L_{1}-\Biggl({{\dot x^{+}}\over{\dot x^{-}_{a}}}L_{2}\Biggr)_{r}=R^{-}
\end{array}\right.$$
$$\Updownarrow$$
\begin{equation} \quad\quad\quad\left\{ \begin{array}{l}
\dot x^{+}L_{2}-(\dot x^{+}L_{2})_{r}=\dot x^{+}R^{+}+\dot x^{-}R^{-}\\
L_{1}={{\dot x^{+}}\over{\dot x^{-}}}(L_{2}-R^{+})
\end{array}\right.\label{ltwo}\end{equation}}
\newcommand{\force}
{\begin{eqnarray}
m_{1}{{d}\over{dx_{0}}}{{v_{x}}\over{\sqrt{1-v_{x}^{2}}}}&=&e_{1}E_{x}
+\Biggl({{d\tau}\over{ds_{x}}}\Biggr)^{2}H(\tau)\label{force}\\
m_{2}{{d}\over{dx_{0}}}{{v_{y}}\over{\sqrt{1-v_{y}^{2}}}}&=&-e_{2}E_{y}
-\Biggl({{d\tau}\over{ds_{y}}}\Biggr)^{2}H(\tau)\nn\\
(\ ds^{2}_{x}=dx_{0}^{2}-dx_{1}^{2},&& ds^{2}_{y}=dy_{0}^{2}-dy_{1}^{2} \ )\nn
\end{eqnarray}}
\newcommand{\geom}
{\begin{eqnarray}
\Delta t'&=&\Delta t\ {{1-v}\over{1+v}}\nn\\
\Delta x_{0}&=&{{\Delta t}\over{1+v}},\quad
\Delta x_{1}={{\Delta t\ v}\over{1+v}}\nn\\
\Delta s^{2}&=&\Delta t^{2}\ {{1-v}\over{1+v}}\nn
\end{eqnarray}}
\newcommand{\mirrow}
{\begin{equation}
k'\Delta t' = k\Delta t,\quad {{\Delta p}\over{\Delta x_{0}}}=
{{2k\Delta t}\over{\Delta s^{2}}} \label{mirrow}
\end{equation}}
\newcommand{\imp}
{\begin{eqnarray}
-m_{1}\Delta n_{x}^{+}(\tau_{i}+n)&=&m_{2}\Delta n_{y}^{+}(\tau_{i}+n-1)
=k_{n}^{+}\label{imp}\\
m_{1}\Delta n_{x}^{-}(\tau_{i}+n)&=&-m_{2}\Delta n_{y}^{-}(\tau_{i}+n)
=k_{n}^{-}\nn
\end{eqnarray}}

A conventional framework for the description of electromagnetic interaction is
a field theory. There is less a known approach to electrodynamics as the
theory with action at a distance in space-time. This approach was proposed
by Schwarzschild \cite{Sch} in 1903, was considered by Tetrode \cite{Tet}
and Fokker \cite{Fok} in the 20th and got its net formulation in papers
of Wheeler and Feynman \cite{FW} in 1945 and 1949. In this theory
electromagnetic field is expressed in terms of
coordinates and velocities of charged particles using Maxwell equations.
The field obtained is
substituted into equations of particles motion. An action in terms
of the world lines of particles is offered which reproduces these equations.

A general solution of these equations is unknown. There is no significant
progress in the investigation of this area. The main obstacle is the following.
The action of electrodynamics contains advanced and retarded terms, hence
the equations of motion relate coordinates and velocities at different
instants of time.
 They are differential equations with a deviating argument. A theory
of such equations is still being developed \cite{dif}. As the equations of
electrodynamics are concerned, it is not clear even whether they have
unique solution. Only some particular solutions have been studied \cite{Driv}.

The presence of advanced and retarded terms in action hinders Hamiltonian
mechanics construction. Hamiltonian description is available only for such
systems, whose Lagrangian depends on coordinates and velocities at one
value of the evolution parameter. Hamiltonian formalism is an effective tool
for the solution of Lagrangian equations of motion, and it is also a basis for
canonical quantization. Failure in constructing Hamiltonian description of
action-at-a-distance electrodynamics stimulated a search for an alternative
quantization procedure. Quantum theory with path integrals was proposed
by Feynman in application to the problem involved \cite{Feyn}.

In this article the Hamiltonian formulation for the problem of motion of
two charged particles in the space free from other charges
is presented. Parametrical
invariance of the action is used. This invariance allows ones
to choose any convenient
parametrization of the world lines. In Section~1 such parametrization is
chosen,
that the argument deviation in advanced and retarded terms of the action will
be
constant. Lagrangian equations are obtained for this the action. In Section~2
boundary conditions are chosen for the equations. In Section~3 the action is
rewritten in the form of integrated Lagrangian depending on the coordinates and
velocities at one value of the integration parameter. The
Hamiltonian formulation is found for this mechanics.

Only one-dimensional motion of charges will be considered in this article.
Generalization to greater number of dimensions adds complexity to the algebra,
but the construction scheme might be preserved.

\section{Equations of motion}
1.\ The motion of charged particles $x$ and $y$, interacting via
electromagnetic field, can be described by stationary principle for
the action \cite{FW}:
\sini
(metric $g^{\mu\nu}=\mbox{\ diag\ }(+1,-1,-1,-1)$ is used)\\
Performing one integration in (\ref{sini}), one can cancel $\delta$-function:
\stptm
where $(x(\tau)-y(\tau^{\pm}(\tau)))^{2}=0,\ \tau^{+}>\tau^{-}$.
Dot denotes differentiation with respect to $\tau$. The action is
parametrically
invariant.

Functions $\tau^{\pm}(\tau)$ mark on the world line of the
particle $y$ the points
of its intersection by a light cone with an origin placed in the point
$x(\tau)$. They are moments when retarded and advanced fields from the point
$x(\tau)$ reach the world line $y$. The $\tau^{\pm}(\tau)$ are monotonically
increasing functions of $\tau$.

\ \\
2.\ There is a parametrization of the
world lines satisfying the following condition:
$$\tau^{+}(\tau)=\tau,\quad \tau^{-}(\tau)=\tau -1$$

Let the world lines be given. Let us consider a light ray,
emitted into the future
from an arbitrary point on the world line $x$, until its
intersection with the world line
$y$. From the point of intersection we draw a ray to the intersection with the
line $x$. Let us continue this light stairway (fig.1).
We will mark the points of reflection
of the light ray from the world lines by the values of parameter
$\tau$ with a unit
step. We specify an arbitrary parametrization $\tau\in [0,1]$ on the segment
of the world line $x$ between two sequential reflections, then define
parametrization of the whole world line using process described above
( shifting the stairway along the segment with initial parametrization~).
We note that parametrization constructed is continuous. We can choose
parametrization on the segment $[0,1]$ so that parametrization will be smooth,
double differentiable and so on.
Continuous parametrization is sufficient for our purpose.

Let us consider one-dimensional motion caused by restricting the initial data
to a straight line. We introduce light coordinates
$x^{\pm}=x_{0}\pm x_{1}$ in a space-time plane containing the world lines.
In the parametrization constructed the following relations hold:
\yx
-- position $y$ is determined by parametrization of the world line $x$.

$x^{\pm},\ y^{\pm}$ are monotonically increasing functions of $\tau$:
$$\dot x^{+}\dot x^{-}=\dot x^{2}>0,\quad
{{1}\over{2}}(\dot x^{+}+\dot x^{-})=\dot x_{0}>0 \ \Longrightarrow
\ \dot x^{\pm}>0 $$

3.\ Let us substitute expressions (\ref{yx}) into the action (\ref{stptm}):
\sr
We denote\qquad $x_{a}(\tau)=x(\tau +1),\ x_{r}(\tau)=x(\tau  -1)$.

The action (\ref{sr}) contains advanced terms only ( shifting the parameter
$\tau\to\tau -1$ in (\ref{sr}), one can obtain the action containing retarded
terms only~). The stationary principle for this action leads to equations:
\eqlag
The equations contain both retarded and advanced terms. Under the action
variation the terms of the form $F\delta x_{a}$ are transformed into the form
$F_{r}\delta x$ via the shift of integration parameter. In this way retarded
terms appear in equations of motion.

Equations (\ref{eqlag}) are still complicated to solve them immediately.

\section{Boundary conditions}
In this Section we show that equations of motion (\ref{eqlag}) have a more wide
class of solutions than equations (\ref{eqmin}), obtained from the same
action in parametrization \  $\tau =$ Minkowski time. Additional solutions are
excluded by imposing proper boundary conditions.

For additional solutions an extra force appears in the right
hand side of the equations
of motion. We show that this force is determined by the type of boundary
conditions rather than by the stationary action principle.
For the problem in question
the extra force is an artifact of parametrization used, it looks like physical
phenomenon. We consider in detail how this force appears.

\ \\
The stationary principle for the
action (\ref{sini}) in parametrization $\tau =x_{0}$
leads to equations \cite{Driv}:
\eqmin
Let us transform these equations in the stairway parametrization:
\eqstar
It is easy to check by substitution that equations (\ref{eqlag})
{\it follow} from these equations : any solution of (\ref{eqstar}) is a
solution
of (\ref{eqlag}).

The converse is incorrect, (\ref{eqlag}) are equivalent to the equations of a
more general form :
\eqls
We denote the left (right) hand side of equations (\ref{eqstar}) by
$L_{1,2}\ (R_{1,2})$.  $H(\tau)$ is an arbitrary function with a period 1.

Proof.
\begin{quotation}
\pro
The first equation in (\ref{ltwo}) is a linear difference equation for
$\dot x^{+}L_{2}$. A particular solution of this equation is
 $$L_{2}=R_{2}.$$
The general solution of this equation has the form:
$$\quad\dot x^{+}L_{2}=\dot x^{+}R_{2}+H\ ,$$
where $H$ is a general solution of the homogeneous equation\\ $H-H_{r}=0$, so
$H$ is a periodical function with period~1. Substituting this solution into
second equation (\ref{ltwo}), we have (\ref{eqls}).
\end{quotation}
Let us transform equations (\ref{eqls}) into parametrization $\tau =x_{0}$:
\force
A new force appears in the right hand side of equations of motion. The
force acts
on segments of the world lines intercepted by two rays of the
light stairway with the
initial values of the parameter $\tau$ and $\tau +d\tau$ (fig.2). The force is
inversely proportional to the square of the segment interval, the
proportionality coefficient $d\tau^{2}H(\tau)$ is common for all segments.

The force of an identical form appears in the other problem. The same force
acts
between two mirrors, moving along a straight line, when radiation is placed
between them.

Let us consider a light impulse of energy $k$ and duration $\Delta t$,
reflected from a moving mirror. From geometrical reasoning (fig.3) we
have:
\geom
Conservation of momentum leads to relations:
$$k'=k\ {{1+v}\over{1-v}},\quad \Delta p=|p'-p|={{2k}\over{1-v}}\quad
\mbox{ ( for small $k$ )}$$
Therefore,
\mirrow
The force (\ref{mirrow}) has the same form as the extra term in (\ref{force}).
The coefficient $k\Delta t$, analogous to $d\tau^{2}H(\tau)$ in (\ref{force}),
conserves in reflections. The analogy is possible for $H(\tau)>0$,
in the case of repulsion.

The extra force essentially affects the structure of the
solution of equations of motion.
The arbitrary function $H(\tau)$ determining the energy distribution for
the radiation between the mirrors brings about the functional ambiguity in
the solution.
This ambiguity remains both in the
non-relativistic limit and the limit $e\to 0$.
(~The mirrors repel even though they move slowly or they are not charged. )

The radiation between the mirrors vanishes at $t\to\infty$. In reflections from
the mirrors,
moving in opposite directions with asymptotically constant velocity, the
duration of light impulse exponentially increases:
$\Delta t_{n}\sim\Delta t_{0}\Bigl({{1+v}\over{1-v}}\Bigr)^{n}\to\infty $
(see fig.2), hence $k\to 0$. The radiation vanishes at $t\to -\infty$ also,
owing to the time symmetry of the problem. In other words, the initial impulse,
as small as one likes is amplified in reflections, causes the finite effect of
repulsion, and then it is attenuated down to zero.

Nevertheless, in Wheeler-Feynman electrodynamics the whole field of radiation
is
expressed in terms of the sources from the
Maxwell equations using conditions of
absence of radiation at infinity, and it is already presented by retarded and
advanced terms in equation (\ref{eqmin}). The new force has no relation to
electrodynamics.

We ask, how the replacement of the parameter in parametrically invariant action
changes the equations of motion. Actions calculated in different
parametrizations coincide. Certainly, the calculation must be performed for one
and the same segment of the trajectory.
When parametrization changes, one must take
care that the initial and final points of the
trajectory do not change. The values of the
parameter, marking boundary points, are obtained from an equation
$x(\tau^{*})=x^{*}$, and they can become variables depending on the trajectory.
Simple example : $S=\int\limits_{0}^{\tau^{*}}d\tau\ \sqrt{\dot x^{2}}=
 \int\limits_{0}^{s^{*}}ds\ ,\quad s^{*}$ must be a dynamical variable for
substantial equations to be obtained.

In the action variation the initial and final points of the
trajectories are fixed~:
$\delta x(\tau_{i})=\delta x(\tau_{f})=\delta y(\tau_{i})=\delta
y(\tau_{f})=0$.
In the
stairway parametrization dynamical variables are the
coordinates of particle $x$,
the information about the trajectory $y$ is ``encoded'' in the
parametrization of  the
trajectory $x$. When the trajectory $x$ is varied so that the initial and final
positions $x$ are fixed than the boundary points of $y$ change freely (fig.4).
Variations of $x$ should be restricted to such class, that boundary points
of $y$ would be fixed. When minimum of the action is sought in this class,
the extra force is excluded.

We describe this calculation, omitting details. The initial and final values of
parameter $\tau_{i}$ and $\tau_{f}$ become dynamical variables. Conditions,
fixing the positions of the
boundary points, are included in the action with Lagrangian
multipliers. Minimum of such action is found with respect to all dynamical
variables. All off-integral terms appearing in calculation should be taken
into account.

Different terms in the
action are turned on at different instants of time (see fig.4).
In interval $\tau\in (\tau_{f}-1,\tau_{f}]$ only the
term representing length of
the world line $x$ is active. The stationary action principle applied to this
interval leads to equations of motion without additional $H$-term. Periodical
condition on function $H$ excludes this term for all instants of time.

Extra force is determined by boundary conditions. If boundary conditions
different from fixation of initial and final coordinates are imposed, distinct
equations of motion would be obtained. When boundary conditions have the form :
\begin{eqnarray}
&&\mbox{``light emitted from the initial point $x_{i}$, after integer
number of reflections }\nn\\
&&\mbox{from the world lines arrives at the final point $x_{f}$''}\label{Zcond}
\end{eqnarray}
force (\ref{mirrow}) appears in equations. One can derive this force in any
parametrization.

Let us consider infinite trajectories. In variation of the
action off-integral term
${{\delta S}\over{\delta\dot x_{\mu}}}\delta x_{\mu}\Bigr|_{-\infty}^{+\infty}$
appears. One excludes this term by requiring $\delta x(\pm\infty)=0$.
In commonly used parametrizations this requirement can be satisfied, because
variation $\delta x$ can always be made local. In stairway parametrization
local disturbance affects the parametrization of the whole trajectory. Size of
response to disturbance exponentially increases when $t\to\infty$ (fig.2).
Requirement $\delta x(\pm\infty)=0$ can be satisfied, if only those correlated
variations of the
trajectories are taken, for which responses induced are cancelled at
infinity (fig.5). Therefore, equations (\ref{eqlag}) give only conditional
minimum of the
action in the class of trajectories variations, where calculations
in the stairway parametrization are well defined.

An exact minimum is defined as follows
$$F_{x}={{\delta S}\over{\delta x}}\ \Bigr|_{\mbox{{\small in cond.min.}}}=0$$
Variation is performed with respect to all other changes of
trajectories.

In calculation of variational derivative the variables $x$ and $\dot x$
are not independent : $${{\delta \dot x(\tau_{1})}\over{\delta x(\tau_{2})}}=
\delta ' (\tau_{1}-\tau_{2})$$ In variation of the action all terms of form
$F\delta\dot x$ can be transformed into the form $-\dot F\delta x$ using
integration in parts. In this the coefficient at $\delta x$ in $\delta S$
is the required derivative ${{\delta S}\over{\delta x}}$.

This calculation should be done in some other parametrization, where it is
defined. As a result we have
$$F_{x}=\Biggl(m_{1}{{d}\over{dx_{0}}}{{v_{x}}\over{\sqrt{1-v_{x}^{2}}}}-
e_{1}E_{x}\Biggr)\Bigr|_{\mbox{{\small in cond.min.}
}}=\Biggl({{d\tau}\over{ds_{x}}}
\Biggr)^{2}H=0\ ,$$
hence, the extra force is excluded.

A similar mechanism determines an additional force in the problem of free
motion
of mass on a surface. Here equations of motion define conditional minimum
of the action in a class of trajectories lying on the surface. The surface acts
on a mass through a reaction force. This force is equal to a
derivative of an action
${{\delta S}\over{\delta x_{\perp}}}$ with respect to variations of coordinates
leading away the mass from the surface. This force can not be obtained from
the stationary principle
of some action formulated in terms of surface coordinates
only. This analogy is correct with the following refinement : the stairway
parametrization constrains not the paths themselves but their variations in
the vicinity of an arbitrary path (fig.6).

Let us consider the finite trajectories again. The
requirement that $\tau_{i}$ and
$\tau_{f}$ be dynamical variables is not necessary. They can be constant,
but their difference should not be integer. Only when
$\tau_{f}-\tau_{i}\in{\bf Z}$ a restriction of class of trajectories occurs ---
for such trajectories the light stairway has integer number of steps
(\ref{Zcond}). When $\tau_{f}-\tau_{i}\notin{\bf Z}$, positions of boundary
points are not related because of sufficient freedom in choice of
parametrization of trajectories in interval $\tau\in [0,1]$. After any
sufficiently small variation of the trajectories with
$\tau_{f}-\tau_{i}\notin{\bf Z}$ light stairway parametrization can be chosen
on them with the same $\tau_{i}$ and $\tau_{f}$.

Further we restrict ourselves just to the trajectories with
$\tau_{f}-\tau_{i}\in{\bf Z}$. The ends of the
trajectories will be fixed. In this case
the equations of motion determine the conditional minimum of the action in
class
of trajectories satisfying condition (\ref{Zcond}). Light stairway with the
initial value of parameter $\tau_{i}$ divides the trajectories into $N$
segments. The requirement
${{\delta S}\over{\delta x}}\ \Bigr|_{\mbox{{\small in cond.min.}}}=0$
for inner points of the segments is automatically fulfilled : even though the
positions of the boundary points of the segments can be related, there is
nothing to constrain the position of inner points. For ``free''
 motion ( in the limit $e\to 0$ ) the
action reaches the minimum on straight-line segments (fig.7). If condition
(\ref{Zcond}) is not fulfilled for straight lines connecting the ends of the
trajectories, then the straight trajectories do not belong to the class,
in which the variation is performed. The conditional minimum of the
action is reached on a polygonal line.

The momentum of a particle is proportional to unit tangent vector to the world
line $n^{\mu}_{x}=\dot x^{\mu}/\sqrt{\dot x^{2}}$. On the polygonal world
lines the momentum of each particle changes. From the relations of the next
Section one can show that for such trajectories the total momentum conserves
(see (\ref{ttr}))~:
\imp
where $\Delta n^{\mu}(\tau)=n^{\mu}(\tau +o)-n^{\mu}(\tau-o)$ is discontinuity
of the unit tangent vector. The force acting on a particle is singular.
It has form (\ref{mirrow}).

The requirement ${{\delta S}\over{\delta x}}\ \Bigr|_{\mbox{{\small
in cond.min.}}}=0$
for the vertices of the polygonal line represents a condition of smoothness of
trajectories:
\begin{eqnarray}
S&=&-m_{1}\sum_{i=1}^{N}\sqrt{(x_{i+1}-x_{i})^{2}}
    -m_{2}\sum_{i=1}^{N-1}\sqrt{(y_{i+1}-y_{i})^{2}}\nn\\
{{\partial S}\over{\partial x_{i}}}&=&m_{1}(\Delta n_{x})_{i}=0\nn
\end{eqnarray}
This condition can be satisfied, only if the exact minimum of the
action belongs to
the class of the allowed variations. It implies that the position of the
ends of the trajectories ( fixed in variations ) is chosen so that the
solution obeys condition (\ref{Zcond}).

By this means, the condition of smoothness of trajectories selects physical
solutions among the trajectories with $\tau_{f}-\tau_{i}\in{\bf Z}$,
minimizing an action. For continuity of the unit tangential
vector in 2-dimensional
space it is sufficient to require the continuity for one of its components.
By virtue of (\ref{imp}), it is sufficient to require continuous linking
of tangent in single point $\tau_{i}+n$.

\def\tz{\begin{equation}
x_{n}^{\pm}(1)=x_{n+1}^{\pm}(0) \label{tz}
\end{equation}}
\def\ton{\begin{eqnarray}
S&=&\int\limits_{0}^{1}d\tau\ \sum_{n}\Bigl(
-m_{1}\sqrt{\dot x_{n}^{+}\dot x_{n}^{-}}
-m_{2}\sqrt{\dot x_{n}^{+}\dot x_{n+1}^{-}}\nn\\
&&-e_{1}e_{2}\Biggl( {{\dot x_{n}^{-}}\over{x_{n+1}^{-}-x_{n}^{-}}} +
{{\dot x_{n}^{+}}\over{x_{n+1}^{+}-x_{n}^{+}}}
\Biggr)\Bigr)\label{ton}
\end{eqnarray}}
\def\ttw{\begin{eqnarray}
p_{n}^{+}&=&-{{m_{1}}\over{2}}\sqrt{{{\dot x_{n}^{-^{\ }}}
\over{\dot x_{n}^{+}}}}
-{{m_{2}}\over{2}}\sqrt{{{\dot x_{n+1}^{-}}\over{\dot x_{n}^{+}}}}
-{{e_{1}e_{2}}\over{x_{n+1}^{+}-x_{n}^{+}}}\quad\quad n=2..N \label{ttw}\\
p_{n}^{-}&=&-{{m_{1}}\over{2}}\sqrt{{{\dot x_{n}^{+^{\ }}}
\over{\dot x_{n}^{-}}}}
-{{m_{2}}\over{2}}\sqrt{{{\dot x_{n-1}^{+}}\over{\dot x_{n}^{-}}}}
-{{e_{1}e_{2}}\over{x_{n+1}^{-}-x_{n}^{-}}}\quad\quad n=2..N \nn\\
\ &&\ \nn\\
p_{1}^{+}&=&-{{m_{1}}\over{2}}\sqrt{{{\dot x_{1}^{-}}\over{\dot x_{1}^{+}}}}
-{{m_{2}}\over{2}}\sqrt{{{\dot x_{2}^{-}}\over{\dot x_{1}^{+}}}}
-{{e_{1}e_{2}}\over{x_{2}^{+}-x_{1}^{+}}}\quad \quad
p_{N}^{+}=-{{m_{1}}\over{2}}\sqrt{{{\dot x_{N}^{-}}\over{\dot x_{N}^{+}}}}
\nn\\
p_{1}^{-}&=&-{{m_{1}}\over{2}}\sqrt{{{\dot x_{1}^{+}}\over{\dot x_{1}^{-}}}}
-{{e_{1}e_{2}}\over{x_{2}^{-}-x_{1}^{-}}}\quad \quad
p_{N}^{-}=-{{m_{1}}\over{2}}\sqrt{{{\dot x_{N}^{+}}\over{\dot x_{N}^{-}}}}
-{{m_{2}}\over{2}}\sqrt{{{\dot x_{N-1}^{+}}\over{\dot x_{N}^{-}}}}\nn
\end{eqnarray}}
\def\tpl{\begin{equation}
\delta S=\int\limits_{0}^{1}d\tau\ \sum_{n}\Biggl(-\dot p_{n}^{\mu}
+{{\partial L}\over{\partial x_{n}^{\mu}}}\Biggr)\delta x_{n}^{\mu}\ +\
\sum_{n}p_{n}^{\mu}\delta x_{n}^{\mu}\Bigl|_{0}^{1}=0\label{tpl}
\end{equation}}
\def\fixed{\begin{equation}\begin{array}{lll}
\delta x_{1}^{+}(0)=\delta y_{1}^{+}(0)=0&
\delta x_{1}^{-}(0)=0&
\delta y_{1}^{-}(0)=\delta x_{2}^{-}(0)=0\\
\delta x_{N}^{-}(1)=\delta y_{N-1}^{-}(1)=0&
\delta x_{N}^{+}(0)=0&
\delta y_{N-1}^{+}(1)=\delta x_{N-1}^{+}(1)=0
\end{array}\label{fixed}\end{equation}}
\def\ttr{\begin{eqnarray}
&&p_{n}^{\mu}(1)=p_{n+1}^{\mu}(0)\quad\quad n=2..N-2\quad
 \mu =+,-\ ;\label{ttr}\\
&&\qquad\qquad n=1\quad \mu =+\ ;\quad n=N-1\quad \mu =-\nn
\end{eqnarray}}
\def\tst{\begin{eqnarray}
r_{n}^{\pm}={{2}\over{m_{1}}}\Biggl(p_{n}^{\pm}+
{{e_{1}e_{2}}\over{x_{n+1}^{\pm}-x_{n}^{\pm}}}\Biggr)\quad n=1..N-1\ ,\quad
r_{N}^{\pm}={{2}\over{m_{1}}}p_{N}^{\pm}\label{tst}\\
\beta ={{m_{1}}\over{m_{2}}}\ ,\quad
v_{n}=\sqrt{{{\dot x_{n}^{-}}\over{\dot x_{n}^{+}}}}\quad n=1..N\ ,\quad
u_{n}=\sqrt{{{\dot x_{n+1}^{-}}\over{\dot x_{n}^{+}}}}\quad n=1..N-1\nn
\end{eqnarray}}
\def\tfo{\begin{eqnarray}
r_{n}^{+}&=& -v_{n}-u_{n}/\beta\quad\quad n=1..N-1\label{tfdiez}\\
r_{n}^{-}&=& -{{1}\over{v_{n}}}-{{1}\over{\beta u_{n}}}\quad\quad n=2..N\nn\\
r_{N}^{+}&=& -v_{N}\qquad\qquad r_{1}^{-}= -{{1}\over{v_{1}}}\nn\\
&&\mbox{\ ($2N$ equations on $2N-1$ variables $v_{n},u_{n}$)}\nn\\
&&\quad\Updownarrow\nn\\
v_{n}&=&-u_{n}/\beta-r_{n}^{+}\quad\quad n=2..N-1\label{tfa}\\
u_{n}&=&{{\beta (r_{n}^{+}r_{n}^{-}-1)\cdot u_{n-1}\ +\ r_{n}^{+}}\over
{\ \quad -r_{n}^{-}\cdot u_{n-1}\ -\ 1/\beta}}\quad\quad n=2..N-1\label{tfb}\\
v_{1}&=&-{{1}\over{r_{1}^{-}}}\qquad u_{1}={{\beta (r_{1}^{+}r_{1}^{-}-1)}
\over{-r_{1}^{-}}}\label{tfc}\\
v_{N}&=&-r_{N}^{+}\qquad u_{N-1}={{-r_{N}^{+}}\over
{\beta (r_{N}^{+}r_{N}^{-}-1)}}\label{tfd}
\end{eqnarray}}
\def\tfi{\begin{eqnarray}
&&u_{n}={{\Psi_{n}^{1}}\over{\Psi_{n}^{2}}}\ ,\
\left(\begin{array}{c}\Psi_{n}^{1}\\ \Psi_{n}^{2}\end{array}\right) =
g_{n}
\left(\begin{array}{c}\Psi_{n-1}^{1}\\ \Psi_{n-1}^{2}\end{array}\right)\ ,
\ g_{n}=
\left(\begin{array}{cc}
\beta (r_{n}^{+}r_{n}^{-}-1)& r_{n}^{+}\\
-r_{n}^{-}& -1/\beta
\end{array}\right) \label{tfi}
\end{eqnarray}}
\def\univ{\left(\begin{array}{c}1\\ 0\end{array}\right)}
\def\unig{\bigl( 1\ 0\bigr)}
\def\inuv{\left(\begin{array}{c}0\\ 1\end{array}\right)}
\def\inug{\bigl( 0\ 1\bigr)}
\def\asv{\left(\begin{array}{c}u_{\pm\infty}\\ 1\end{array}\right)}
\def\tsi{\begin{equation}
\Delta =\unig\ g_{N}...g_{1}\ \univ =0\label{tsi}
\end{equation}}
\def\tsip{\begin{equation}
\Delta =\unig\ g_{N}...g_{-N}\ \univ =0\label{tsip}
\end{equation}}
\def\tsipp{\begin{equation}
\Delta =\unig\ g_{+\infty}^{(0)}\ g_{+\infty}...g_{1}\ g_{0}\ g_{-1}...
g_{-\infty}\ g_{-\infty}^{(0)} \univ =0\label{tsipp}
\end{equation}}
\def\asz{\begin{eqnarray}
r_{-N}^{-}\to -{{1}\over{v_{-\infty}}}\ ,&&\quad r_{N}^{+}\to -v_{+\infty}\quad
\quad N\to\infty\label{asz}\\
r_{N}^{-}\to -{{1}\over{v_{+\infty}}}-{{1}\over{\beta u_{+\infty}}}\ ,&&\quad
r_{-N}^{+}\to -v_{-\infty}-u_{-\infty}/\beta\nn
\end{eqnarray}}
\def\asy{\begin{eqnarray}
&&r_{n}^{-}\to -{{1}\over{v_{\pm\infty}}}-{{1}\over{\beta u_{\pm\infty}}}\ ,
\quad r_{n}^{+}\to -v_{\pm\infty}-u_{\pm\infty}/\beta\label{asy}\\
&&|n|<N\ ,\quad n\to\pm\infty\ ,\quad N\to\infty\ ,\nn
\end{eqnarray}}
\def\cheig{$$
\left(\begin{array}{cc}
{{1}\over{\beta}}+{{u_{\pm\infty}}\over{v_{\pm\infty}}}+
{{v_{\pm\infty}}\over{u_{\pm\infty}}}&-{{1}\over{\beta}}u_{\pm\infty}
-v_{\pm\infty}\\
{{1}\over{\beta u_{\pm\infty}}}+{{1}\over{v_{\pm\infty}}}&-{{1}\over{\beta}}
\end{array}\right)
\asv = {{u_{\pm\infty}}\over{v_{\pm\infty}}}\asv
$$}
\def\eigv{\begin{eqnarray}
g_{-\infty}^{(0)}\univ &=&{{1}\over{v_{-\infty}}}
\left(\begin{array}{c}u_{-\infty}\\ 1\end{array}\right)
\sim \Psi_{-\infty}\nn\\
\unig\ g_{+\infty}^{(0)} &=&{{v_{+\infty}}\over{u_{+\infty}}}
\ \bigl(1\ \ -u_{+\infty}\bigr)\sim \tilde\Psi_{+\infty}\nn
\end{eqnarray}}
\def\drpx{\begin{eqnarray}
{{\partial}\over{\partial p_{n}^{\pm}}}\Delta(r)&=&
{{2}\over{m_{1}}}{{\partial}\over{\partial r_{n}^{\pm}}}\Delta(r)\nn\\
{{\partial}\over{\partial x_{n}^{\pm}}}\Delta(r)&=&
{{2e_{1}e_{2}}\over{m_{1}}}\Biggl({{1}\over{\bigl(
x_{n+1}^{\pm}-x_{n}^{\pm}\bigr)^{2}}}{{\partial}\over{\partial r_{n}^{\pm}}}
-{{1}\over{\bigl(x_{n}^{\pm}-x_{n-1}^{\pm}\bigr)^{2}}}
{{\partial}\over{\partial r_{n-1}^{\pm}}}\Biggr)\Delta(r)\quad
\quad\mbox{see}\footnotemark\nn
\end{eqnarray}
\footnotetext{When $n=N$, the first item is absent inside brackets,
when $n=1$, the second item is absent}
\newcounter{tmp}
\setcounter{tmp}{\value{footnote}}}
\def\rpnt{\begin{eqnarray}
\dot r_{n}^{\pm}&=&{{2}\over{m_{1}}}\Biggl(\dot p_{n}^{\pm} - e_{1}e_{2}
\ {{\dot x_{n+1}^{\pm}-\dot x_{n}^{\pm}}
\over{\bigl(x_{n+1}^{\pm}-x_{n}^{\pm}\bigr)^{2}}}\Biggr)\quad
\quad\mbox{see}\footnotemark\nn\\
&&=-{{4\lambda e_{1}e_{2}}\over{m_{1}^{2}}}\Biggl({{1}\over{\bigl(
x_{n+1}^{\pm}-x_{n}^{\pm}\bigr)^{2}}}{{\partial}\over{\partial r_{n+1}^{\pm}}}
-{{1}\over{\bigl(x_{n}^{\pm}-x_{n-1}^{\pm}\bigr)^{2}}}
{{\partial}\over{\partial r_{n-1}^{\pm}}}\Biggr)\Delta(r)\qquad
\qquad\mbox{see}\footnotemark[\value{tmp}]\nn
\end{eqnarray}
\footnotetext{When $n=N$, the second item is absent}}
\def\dr{\begin{eqnarray}
{{\partial g_{n}}\over{\partial r_{n}^{+}}}&=&
\left(\begin{array}{cc}
\beta r_{n}^{-}&1\\ 0&0
\end{array}\right)=-\beta\univ\inug\  g_{n}\nn\\
{{\partial g_{n}}\over{\partial r_{n}^{-}}}&=&
\left(\begin{array}{cc}
\beta r_{n}^{+}&0\\ -1&0
\end{array}\right)=\beta\ g_{n}\inuv\unig \qquad
{{\partial g_{k}}\over{\partial r_{n}^{\pm}}}=0\ ,\ k\ne n \nn\\
\Delta_{n}^{+}&=&{{\partial \Delta}\over{\partial r_{n}^{+}}}=
-\beta\cdot\unig\ g_{N}...g_{n+1}\univ\cdot\inug\ g_{n}...g_{1}\univ \nn\\
\Delta_{n}^{-}&=&{{\partial \Delta}\over{\partial r_{n}^{-}}}=
\beta\cdot\unig\ g_{N}...g_{n}\inuv\cdot\unig\ g_{n-1}...g_{1}\univ \nn\\
\Delta_{n}^{+}&=&-\beta\tilde\Psi_{n}^{1}\Psi_{n}^{2}\label{tte}\\
\Delta_{n}^{-}&=&\beta\tilde\Psi_{n-1}^{2}\Psi_{n-1}^{1}\nn\\
\mbox{where}&&\Psi_{n}=g_{n}...g_{1}\univ\sim
\left(\begin{array}{c}u_{n}\\ 1\end{array}\right)\ ,\nn\\
&&\tilde\Psi_{n}=\unig\ g_{N}...g_{n+1}\sim
\bigl(1\ \ -u_{n}\bigr)\label{teip}\\
&&\Delta =\tilde\Psi_{n}^{1}\Psi_{n}^{1}+\tilde\Psi_{n}^{2}\Psi_{n}^{2}=0
\label{tei}
\end{eqnarray}}
\def\tni{\begin{equation}\left\{ \begin{array}{l}\displaystyle
\dot x_{n}^{\pm}={{2\lambda}\over{m_{1}}}\ \Delta_{n}^{\pm}(r)\\
\dot r_{n}^{\pm}=-{{4\lambda e_{1}e_{2}}\over{m_{1}^{2}}}\Biggl(
{{\Delta_{n+1}^{\pm}(r)}\over{\bigl(x_{n+1}^{\pm}-x_{n}^{\pm}\bigr)^{2}}}-
{{\Delta_{n-1}^{\pm}(r)}\over{\bigl(x_{n}^{\pm}-x_{n-1}^{\pm}\bigr)^{2}}}
\Biggr)\qquad\quad\mbox{see}\footnotemark[\value{tmp}]\end{array}\right.
\label{tni}\end{equation}}
\def\frs{\begin{eqnarray}
{{\dot x_{n+1}^{-}}\over{\dot x_{n}^{+}}}&\stackrel{(\ref{tni})}{=}&
{{\Delta_{n+1}^{-}}\over{\Delta_{n}^{+}}}\stackrel{(\ref{tte})}{=}
-{{\tilde\Psi_{n}^{2}}\over{\tilde\Psi_{n}^{1}}}\
{{\Psi_{n}^{1}}\over{\Psi_{n}^{2}}}\stackrel{(\ref{tei})}{=}
\Biggl({{\Psi_{n}^{1}}\over{\Psi_{n}^{2}}}\Biggr)^{2}\stackrel{(\ref{tfi})}{=}
(u_{n})^{2}\ \nn\\
{{\dot x_{n}^{-}}\over{\dot x_{n}^{+}}}&\stackrel{(\ref{tni})}{=}&
{{\Delta_{n}^{-}}\over{\Delta_{n}^{+}}}\stackrel{(\ref{tte})}{=}
-{{\tilde\Psi_{n-1}^{2}}\over{\tilde\Psi_{n}^{1}}}\
{{\Psi_{n-1}^{1}}\over{\Psi_{n}^{2}}}{=}
( v_{n})^{2}\quad\mbox{compare with (\ref{tst})}\nn
\end{eqnarray}}
\def\prfrs{\begin{eqnarray}
v_{n}&\stackrel{(\ref{tfa})}{=}&-u_{n}/\beta-r_{n}^{+}\stackrel{(\ref{tfi})}{=}
-\bigl(1\ \ -u_{n}\bigr)\ g_{n}\inuv\stackrel{(\ref{teip})}{=}\nn\\
&=&-{{1}\over{\tilde\Psi_{n}^{1}}}\cdot\tilde\Psi_{n}\ g_{n}\inuv
\stackrel{(\ref{teip})}{=}
-{{\tilde\Psi_{n-1}^{2}}\over{\tilde\Psi_{n}^{1}}}\nn\\
{{\Psi_{n-1}^{1}}\over{\Psi_{n}^{2}}}&=&{{1}\over{\Psi_{n}^{2}}}\cdot
\unig\ g_{n}^{-1}\Psi_{n}\stackrel{(\ref{tfi})}{=}{{1}\over{\Psi_{n}^{2}}}\cdot
\unig\ \left(\begin{array}{cc}
 -1/\beta& -r_{n}^{+}\\
r_{n}^{-}&\beta (r_{n}^{+}r_{n}^{-}-1)
\end{array}\right)\Psi_{n}=\nn\\
&=&-u_{n}/\beta-r_{n}^{+}=v_{n}\nn
\end{eqnarray}}
\def\chD{\begin{eqnarray}
\dot\Delta&=&\sum_{n=1}^{N}\dot r_{n}^{\mu}{{\partial\Delta}\over{\partial
r_{n}^{\mu}}}=\nn\\
&=&-{{4\lambda e_{1}e_{2}}\over{m_{1}^{2}}}\sum_{\mu}\Biggl(
\sum_{n=1}^{N-1}
{{\Delta_{n+1}^{\mu}\Delta_{n}^{\mu}}\over
{\bigl(x_{n+1}^{\mu}-x_{n}^{\mu}\bigr)^{2}}}
-\sum_{n=2}^{N}
{{\Delta_{n-1}^{\mu}\Delta_{n}^{\mu}}\over
{\bigl(x_{n}^{\mu}-x_{n-1}^{\mu}\bigr)^{2}}}
\Biggr)=0\nn
\end{eqnarray}}
\def\cont{\begin{eqnarray}
x_{n}^{\mu}(T)&=&x_{n+1}^{\mu}(0)\quad n=1..N-1\ ,\quad
x_{N}^{\mu}(T)=x_{f}^{\mu}
\label{cont}\\
p_{n}^{\mu}(T)&=&p_{n+1}^{\mu}(0)\quad n=2..N-2\quad \mu =+,-\ ; \nn\\
&&\qquad\qquad n=1\quad \mu =+\ ;\quad n=N-1\quad \mu =-\nn
\end{eqnarray}}
\def\contp{\begin{equation}
u_{n}(T)=u_{n+1}(0)\label{contp}
\end{equation}}
\def\contpp{\begin{equation}
(x,p)_{T}=S\ (x,p)_{0}\label{contpp}
\end{equation}}

\section{Hamiltonian formulation}
1.\ Let us break up the axis of parameter $\tau$ into unit intervals. We denote
the values of the coordinates of particle $x$ in each interval as $x_{n}$:
$$ x_{n}^{\pm}(\tau)=x^{\pm}(\tau +n)\ ,\quad\tau\in [0,1]\ ,
\quad n\in {\bf Z}$$
In view of continuity of function $x(\tau)$, a condition is imposed on
the coordinates $x_{n}$: \tz
Let us rewrite action (\ref{sr}) into a form  \ton
To avoid divergence we will consider finite trajectories: summation in
(\ref{ton}) is performed from $n=1$ to $n=N$. When $n=N$ only the first term
$-m_{1}\sqrt{\dot x_{N}^{+}\dot x_{N}^{-}}$ is present in the action.

The action (\ref{ton}) is given in the form of integral
$S=\int d\tau\ L(x_{n},\dot x_{n})$ of the function depending on the
coordinates and
their first derivatives at one value of parameter $\tau$. Hamiltonian
description is available for this mechanics.

\ \\
2.\ Let us define momenta
$p_{n}^{\mu}={{\partial L}\over{\partial\dot x_{n}^{\mu}}}$, conjugated to
the coordinates $x_{n}^{\mu}$~: \ttw
The stationary principle of action (\ref{ton}) has a form  \tpl
An equality to zero of the first item manifests a minimum of the action with
respect
to those variations of the trajectories, when points $x_{n}(0),\ x_{n}(1)$
are fixed. This condition leads to equation (\ref{eqlag}). The second item in
(\ref{tpl}) represents off-integral terms extracted in variation of the action
(\ref{ton}). An equality to zero of the second item is a condition of
the action
minimum with respect to variations of positions $x_{n}(0),\ x_{n}(1)$.
The boundary points $x_{1}(0),\ y_{1}(0)$ and $x_{N}(1),\ y_{N-1}(1)$
are fixed in the variations: \fixed
With regard to these conditions and the condition of continuous linking of
coordinates (\ref{tz}) the requirement of the
equality to zero of the second item
in (\ref{tpl}) is written as condition of continuous linking of the momenta
\ttr

\ \\
3.\ The action (\ref{ton}) is parametrically invariant, hence Legendre
transformation (\ref{ttw}) is degenerate: momenta $p_{n}^{\mu}$ do not change
in replacing $\dot x_{n}^{\mu}\to\lambda\dot x_{n}^{\mu}$. This implies that
the momenta can not be independent, there is a relation
among them. This relation (Hamiltonian constraint) has
a sense of compatibility condition of system (\ref{ttw}) for the velocities
$\dot x_{n}$. Let us obtain this condition.

We introduce notations  \tst
Variables $v_{n}, u_{n}$ are sufficient to determine the velocities:
$$\sqrt{\dot x_{n}^{-}}= v_{n}\sqrt{\dot x_{n}^{+}}\qquad
\sqrt{\dot x_{n}^{+}}={{u_{n-1}}\over{v_{n}}}\sqrt{\dot x_{n-1}^{+}}=
\prod\limits_{k=2}^{n}{{u_{k-1}}\over{v_{k}}}  \cdot\sqrt{\dot x_{1}^{+}}$$
value $\dot x_{1}^{+}$ is arbitrary.

Let us rewrite equations (\ref{ttw}): \tfo
Relation (\ref{tfa}) expresses $v_{n}$ in terms of $u_{n}$. Sequential values
$u_{n}$ are related via linear-fractional transformation (\ref{tfb}).
Let us represent this transformation in a matrix form:  \tfi
The initial value $u_{1}$
for recurrent formula (\ref{tfb}) is given by expression
(\ref{tfc}). In a matrix form:
$$ u_{1}={{\Psi_{1}^{1}}\over{\Psi_{1}^{2}}}\quad \Psi_{1}=g_{1}\univ\
\Longrightarrow\ \Psi_{n}=g_{n}..g_{1}\univ$$
Condition (\ref{tfd}) for value $u_{N-1}$ can be written as
$$\beta (r_{N}^{+}r_{N}^{-}-1)\Psi_{N-1}^{1}\ +\ r_{N}^{+}\Psi_{N-1}^{2}=0
\ \Longleftrightarrow\ \unig g_{N}\Psi_{N-1}=0$$
or \tsi
Equation (\ref{tsi}) is a desired condition on $(x_{n},p_{n})$,
when system (\ref{ttw}) is compatible.

\ \\

Let us consider a limit of infinite trajectories. We change the numeration of
the variables: let $n$ changes from $-N$ to $N$, in the limit $N\to\infty$
the initial and final points $x_{-N},x_{N}$ tend to infinity.
The constraint is \tsip
Variables $v_{n}, u_{n}$ determine the slopes of tangents to the
world lines $x$ and
$y$ (see (\ref{tst})). They have definite limits at $n\to\infty$:
$$\lim_{n\to\pm\infty} v_{n}=v_{\pm\infty}\quad
\lim_{n\to\pm\infty} u_{n}=u_{\pm\infty}$$
Variables $r_{n}^{\mu}$ have definite limits also. In view of (\ref{tfdiez}),
values $r_{n}^{\mu}$ at the
initial and final segments of trajectories are defined in a
special way to compared with
other $r_{n}^{\mu}$. The limiting values \asz differ from those
for inner segments \asy the associated matrices are different too:
$$ g_{\pm\infty}=\lim_{N\to\infty} g_{\pm (N-1)}\ ,\quad
g^{(0)}_{\pm\infty}=\lim_{N\to\infty} g_{\pm N}\ .$$
Asymptotic values $u_{\pm\infty}$ are stationary points of transformation
(\ref{tfb}) at\\ $n\to\infty$. This means that $\Psi_{\pm\infty}=\asv$ are
eigen
vectors of matrices $g_{\pm\infty}$:
\begin{equation}
 g_{\pm\infty}\Psi_{\pm\infty}\sim\Psi_{\pm\infty}\ .\label{eig}
\end{equation}
One can verify this directly, substituting (\ref{asy}) into (\ref{tfi}):
\cheig
Condition (\ref{eig}) can also be written in a form
$$ \tilde\Psi_{\pm\infty}g_{\pm\infty}\sim\tilde\Psi_{\pm\infty}\ ,\quad
\tilde\Psi_{\pm\infty}=\bigl(1\ \ -u_{\pm\infty}\bigr)$$

Proof.
\begin{quotation}
It follows from $g\Psi\sim\Psi$ that $\tilde\Psi g\Psi =0$, where
$\tilde\Psi =\Psi^{T}\left(\begin{array}{cc}0&-1\\1&0\end{array}\right)$.
On the other hand, a general solution of a linear equation on $V$: $\ V\Psi =0$
in 2-dimensional space is $V\sim\tilde\Psi$. Hence $\tilde\Psi
g\sim\tilde\Psi$.
\end{quotation}
Substituting (\ref{asz}) into (\ref{tfi}), one can easily show, that \eigv
Therefore, in the limiting expression of constraint (\ref{tsip}): \tsipp
the facings $\unig\ g_{+\infty}^{(0)}$ and $g_{-\infty}^{(0)}\univ$ are
eigen vectors of the asymptotic matrices $g_{+\infty}$ and $g_{-\infty}$,
respectively. Hence one can insert any number of asymptotic matrices
after left facing and before right facing, and equation (\ref{tsipp})
holds true.

\ \\

The constraint is just a consequence of momenta definition (\ref{ttw}), it is
fulfilled for any world line. We will draw an analogy to mechanics of free
relativistic particle $L=m\sqrt{\dot x^{2}}$: here the constraint
$p^{2}-m^{2}=0$ on momentum $p^{\mu}=m \dot x^{\mu}/\sqrt{\dot x^{2}}$
is fulfilled for any world line. Equations of motion have not been taken into
account yet.

\ \\
4.\ Canonical Hamiltonian for the action (\ref{ton}) vanishes:
$$H_{c}=\sum_{n}p_{n}^{\mu}\dot x_{n}^{\mu} - L =0$$
According to Dirac's description of constrained Hamiltonian systems, the
Hamiltonian is constraint (\ref{tsi}):
$$H=\lambda\Delta\approx 0\ ,$$
$\lambda$ is Lagrangian multiplier ( an arbitrary function of $\tau$ ).
Weak equality symbol implies that condition $\Delta =0$ must be considered
after Poisson brackets calculation.

Hamiltonian equations of motion are
$$\dot x_{n}^{\pm}=\lambda\{ x_{n}^{\pm},\Delta\} =\lambda{{\partial\Delta}
\over{\partial p_{n}^{\pm}}}\ ,\quad
\dot p_{n}^{\pm}=\lambda\{ p_{n}^{\pm},\Delta\} =-\lambda{{\partial\Delta}
\over{\partial x_{n}^{\pm}}}$$
In view of (\ref{tst}), \drpx Hence \rpnt Let us calculate the derivatives
${{\partial\Delta}\over{\partial r}}$ \dr
Hamiltonian equations are \tni

\ \\

Let us derive the ratios \frs In the proof of the last equality the following
identities were used \prfrs
The first equation in (\ref{tni}) is fulfilled for any world line when momenta
definitions in terms of velocities are substituted and a proper Lagrangian
multiplier $\lambda =m_{1}\dot x_{1}^{+}/\Bigl(2\Delta_{1}^{+}(r(\dot
x))\Bigr)$
is chosen. Analogously to mechanics of relativistic particle, where an equation
$\dot x^{\mu}=\lambda\{ x^{\mu},\ p^{2}-m^{2}\} =2\lambda p^{\mu}$
after substitution $p^{\mu}=m{{\dot x^{\mu}}\over{\sqrt{\dot x^{2}}}}$
and choice $\lambda ={{\sqrt{\dot x^{2}}}\over{2m}}$ is satisfied identically
on any world line. The shape of the
world line is determined by the second equation in (\ref{tni}).

Equations (\ref{tni}) conserve the constraint condition:
$\dot\Delta =\lambda\{ \Delta ,\Delta\} =0$. Let us verify this directly: \chD

Hamiltonian equations (\ref{tni}) specify a phase flow on a surface of
constraint (\ref{tsi}), which stratifies this surface into non-intersecting
phase trajectories. The projection of the phase trajectory into configuration
space $x_{n}$ gives the solution of Lagrangian equations for the
action (\ref{ton}).
It is necessary to select from all phase trajectories those, that match
continuous linking conditions.

\ \\
5.\ Lagrangian multiplier in (\ref{tni}) affects only the parametrization of
phase trajectories. It can be excluded by the parameter replacing:
$$\tau\to\tilde\tau(\tau) =\int\limits_{0}^{\tau}\lambda(\tau ')\ d\tau '$$
Value $\tilde\tau(1)=T$ enters in the linking conditions. In new
parametrization
these conditions have the form:  \cont
Values $\Bigl(x_{n}^{\mu},p_{n}^{\mu}\Bigr)_{T}$ are determined in the
solution of
differential equations (\ref{tni}), they are functions of initial data
$\Bigl(x_{n}^{\mu},p_{n}^{\mu}\Bigr)_{0}$ and ``time'' $T$. Conditions
(\ref{cont}) are the system of non-linear equations on initial data and $T$.
Counting shows that the number of equations equals the number of independent
variables (see fig.8):
\begin{eqnarray}
\mbox{$2N$ conditions on coordinates} &+&
\mbox{$2(N-2)$ conditions on momenta +}\nn\\
\mbox{+ $1$ constraint (\ref{tsi}) \quad}&=&
\mbox{\quad $ (2N-4)$ coordinates +}\nn\\
\mbox{ + $2N$ momenta }&+&
\mbox{ $1$ Lagrangian multiplier ( or ``time'' $ô$) }\nn
\end{eqnarray}

In the previous Section it was shown that excluding of non-physical solutions
requires continuous linking of tangent in one point of the trajectory  \contp
To satisfy this condition one should carry some fixed boundary value
(~e.g., $x_{N}^{+}(0)$ ) into the set of independent variables.

\ \\

In the limit $N\to\infty\quad (n=-N..N)\quad$ the sense of the linking
conditions becomes more clear:  \contpp
where a transformation $S\ (x_{n},p_{n})=(x_{n+1},p_{n+1})$ is a shift of
index $n$.

The index shift does not change constraint condition (\ref{tsipp}):
$$\Delta =\tilde\Psi_{+\infty}...g_{n+1}\ g_{n}\ g_{n-1}...\Psi_{-\infty}=0
\stackrel{S}{\to}
\Delta =\tilde\Psi_{+\infty}...g_{n+2}\ g_{n+1}\ g_{n}...\Psi_{-\infty}=0$$
and equations of motion (\ref{tni}) ( because $\Delta_{n}^{\pm}(Sr)=
\Delta_{n+1}^{\pm}(r)\ $ ). This means that transformation $S$ transfers the
phase trajectories into the
phase trajectories. Requirement (\ref{contpp}) implies that
phase trajectories $(x,p)_{\tau}$ and $S\ (x,p)_{\tau}$ should have a common
point. This is possible only when they coincide ( up to reparametrizations ).
Therefore, condition (\ref{contpp}) selects those phase trajectories, that
transform into themselves by transformation $S$.

Let point $(x,p)$ belong to a plane $x_{0}^{+}=C$. Let us apply to it the
transformation $S$, then bring back it onto the plane $x_{0}^{+}=C$ by movement
$R$ along the phase trajectory. The phase trajectory
involved transforms into itself
by transformation $S$ only if   $$ RS\ (x,p)  =(x,p)$$
Therefore, the problem reduces to a search for stationary points of discrete
transformation $RS$.

\ \\

Continuous linking conditions (\ref{cont}),(\ref{contp}) are non-Hamiltonian.
They relate values of dynamical variables at distinct evolution parameters
$\tau$. Mechanics considered has yet another element non-traditional for
Hamiltonian description: parameter $\tau$, used in this Section, is not time
in any reference frame. Coordinates $x_{n}(\tau)$ at one $\tau$ and different
$n$ are separated by time-like interval. Usually the parametrization of
the world line is introduced with the aid of its sectioning by a
specified set of
space-like surfaces \cite{Dir}. This construction is convenient to achieve the
monotonic parametrization. However,
it is not necessary, its rejection does not lead to
contradictions. Let us remember that Hamiltonian description is applied just
to those variational problems, whose formulation does not mention time. So,
Hamiltonian formalism enables one to obtain the shape of minimal surfaces both
in the Minkowski space ( world sheets of relativistic string ) and Euclidean
space ( soap films ).

Nevertheless, just the absence of time interpretation for parameter $\tau$
in our problem allows one to bypass ``no-interaction'' theorem \cite{noint}.
According to this theorem, the following requirements
\begin{enumerate}
\item
single-time Hamiltonian description is applied (~i.e.
one-parametrical Hamiltonian
description, in which the parameter is time in some reference frame~)
\item
coordinates of particles in the Minkowski space are canonical coordinates of
Hamiltonian mechanics
\item
world lines of particles are curves in the Minkowski space, on which Lorentz
group acts by rotations and reparametrizations
\end{enumerate}
are compatible only in mechanics of non-interacting particles. For the
mechanics under discussion requirements 2 and 3 are satisfied. Requirement 1
is violated: canonical coordinates are the set $x_{n}^{\mu}$ of coordinates
of one particle in different moments of the Minkowski time.

\section*{Results}

Hamiltonian mechanics of one-dimensional two body problem in Wheeler-Feynman
electrodynamics is constructed as follows. In phase space $(x_{n}^{\mu},
p_{n}^{\mu})$\\ $n=1..N\quad$ constraint (\ref{tsi}) is given. Hamiltonian
equations (\ref{tni}) specify the phase flow on the surface of the constraint.
Conditions (\ref{cont}),(\ref{contp}) select the physical solutions among
phase trajectories.

So, the structure of solutions of Lagrangian equations --
differential-difference equations (\ref{eqlag}) -- is determined by the system
of ordinary differential equations (\ref{tni}) and the system of non-linear
equations (\ref{cont}),(\ref{contp}). A separate paper will be devoted to
the examination of this structure.

\vspace{1cm}

I would like to thank S.N.Sokolov and G.P.Pron'ko for valuable discussions.

\newpage
\section*{Captions to figures}
\begin{description}
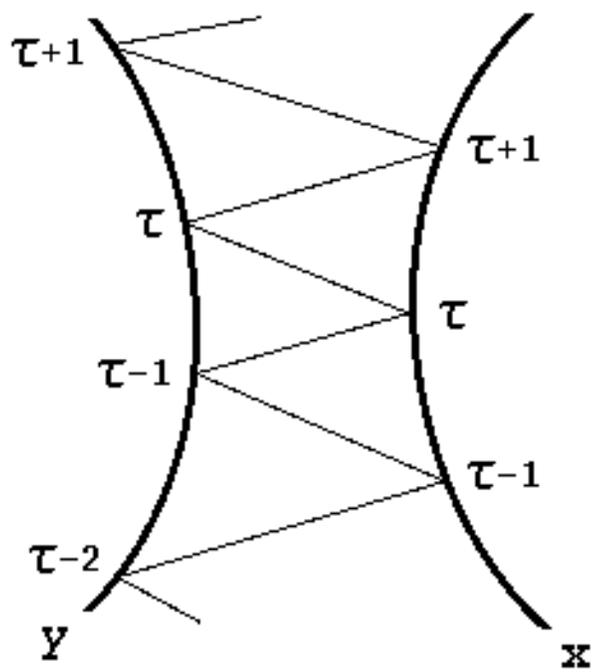
\item fig.1\quad Stairway parametrization.
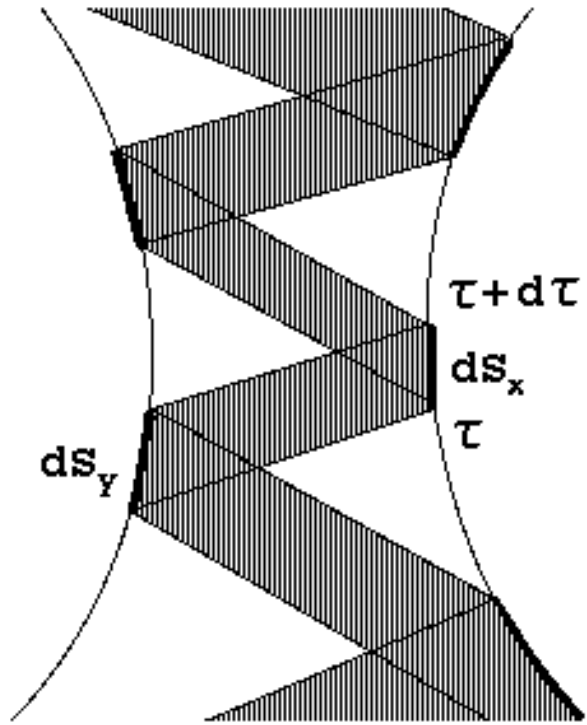
\item fig.2\quad Extra force acts on the marked segments of the world lines.
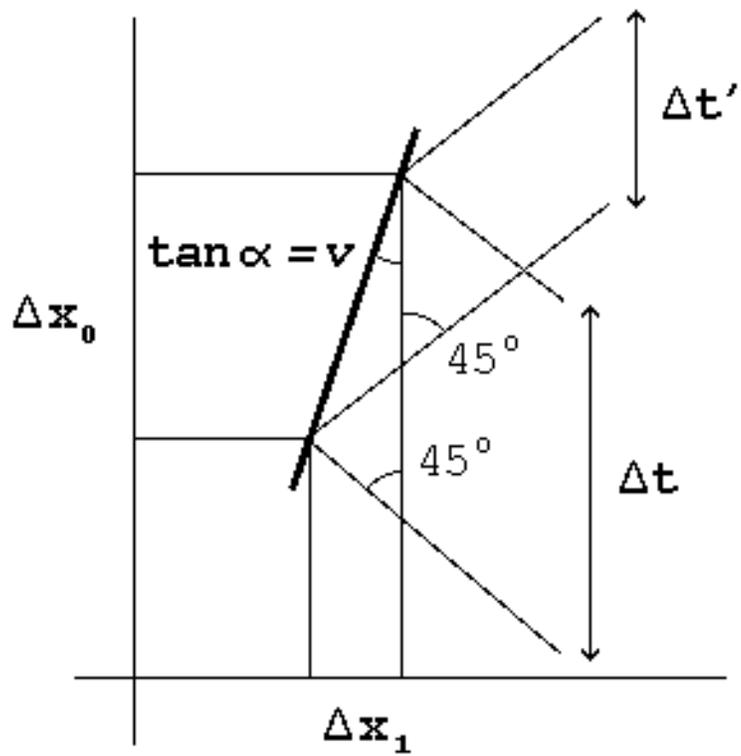
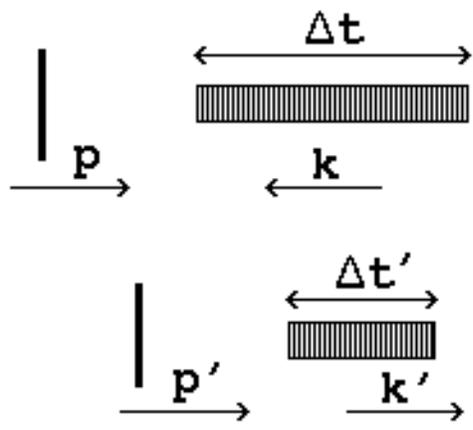
\item fig.3\quad Light impulse reflects from the moving mirror.
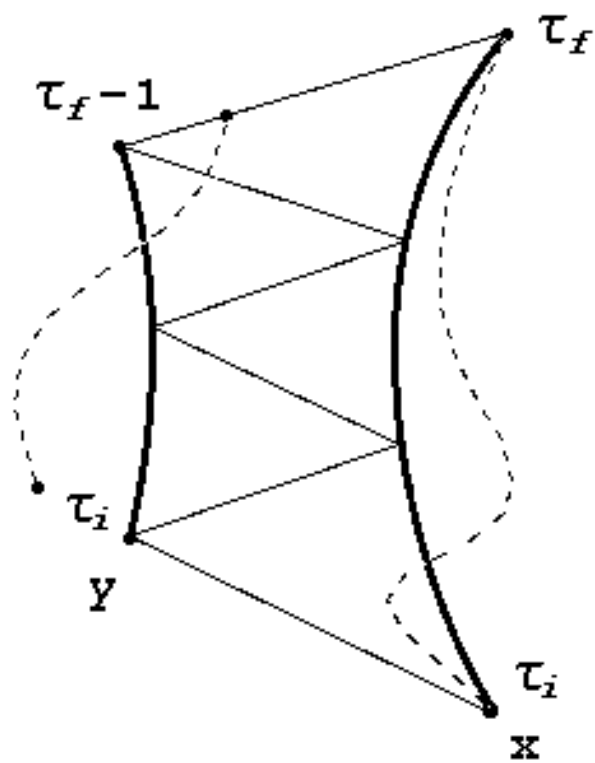
\item fig.4\quad When the boundary points of $x$ are fixed, the boundary
points of $y$ change freely.
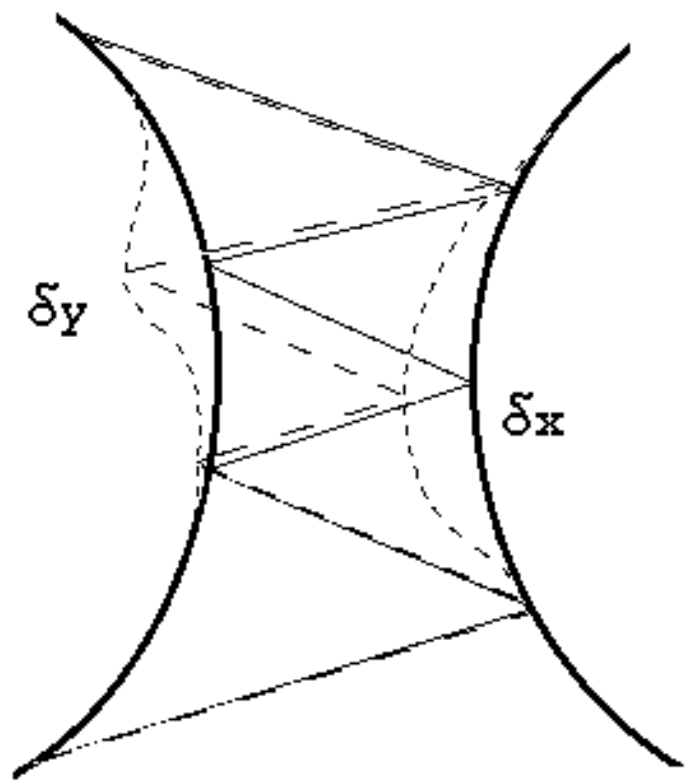
\item fig.5\quad Under the allowed variations the shifts of parametrization
are cancelled at infinity.
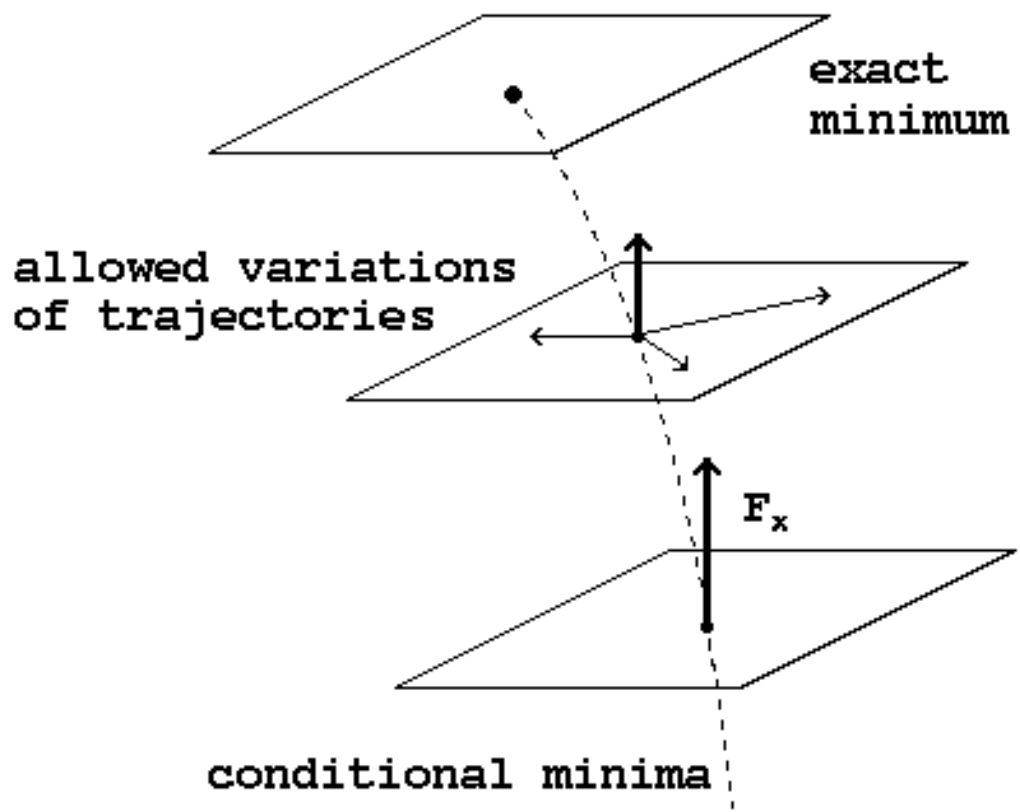
\item fig.6\quad Action minima.
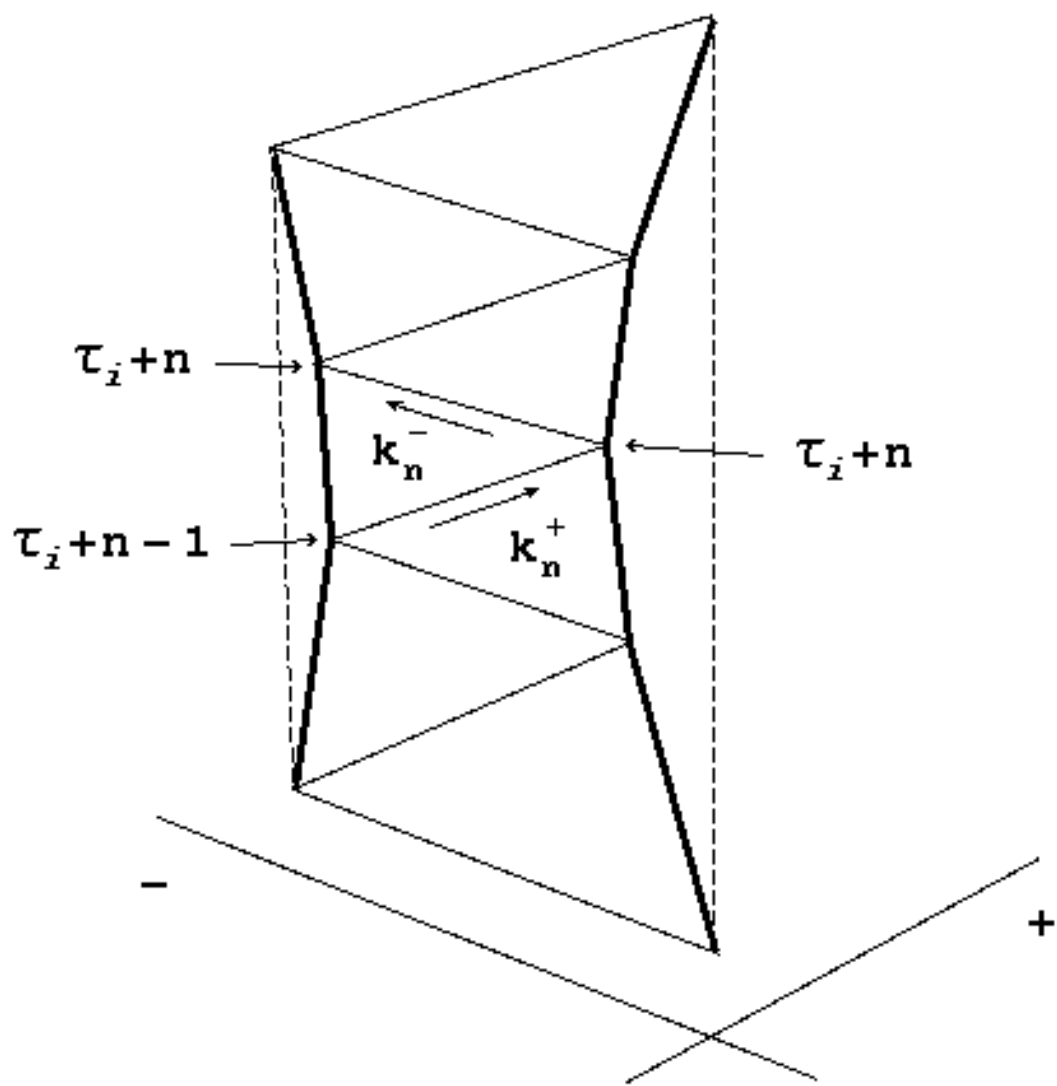
\item fig.7\quad If the light stairway is forced to have an integer number
of steps, the minimum of the action is reached on a polygonal line.
\item fig.8\quad Number of initial data is equal to the number of linking
conditions.
\end{description}

\end{document}